\input phyzzx
\input epsf

\def\scrip{{\cal I^+}}
\def\scrim{{\cal I^-}}
\def\ty{{\tilde y}}
\def\tR{{\tilde R}}
\def\ty{{\tilde y}}
\def\tU{{\tilde U}}
\def\tV{{\tilde V}}
\def\w{{\omega}}

\rightline{UATP-98/02}
\rightline{March, 1998}
\rightline{Revised: September, 1998}
\vskip 0.2in
\centerline{\seventeenbf On the Spectrum of the Radiation}
\centerline{\seventeenbf from a Naked Singularity}
\vskip 0.5in
\centerline{\caps Cenalo Vaz\footnote{\dagger}{Email:
cvaz@mozart.si.ualg.pt}}
\centerline{\it Unidade de Ci\^encias Exactas e Humanas}
\centerline{\it Universidade do Algarve}
\centerline{\it Campus de Gambelas, P-8000 Faro, Portugal}
\vskip 0.1in
\centerline{and}
\vskip 0.1in
\centerline{\caps Louis Witten\footnote{\dagger\dagger}{Email:
witten@physunc.phy.uc.edu}}
\centerline{\it Department of Physics}
\centerline{\it University of Cincinnati}
\centerline{\it Cincinnati, OH 45221-0011, U.S.A.}
\vskip 0.5in
\centerline{\bf \caps Abstract}
\vskip 0.1in

\noindent In the final stages of collapse, quantum radiation due to particle
creation from a naked singularity is expected to be significantly different
from black hole radiation. In certain models of collapse it has been shown
that, neglecting the back reaction of spacetime,  the particle flux on future
null infinity grows as the inverse square of the distance from the Cauchy
horizon. This is to be contrasted with the flux of radiation from a black hole,
which approaches a constant (inversely proportional to the square of
its mass) in the neighborhood of its event horizon. The spectrum of black hole
radiation is identical to that of a black body at temperature $T = (8\pi M
)^{-1}$. We derive the radiation spectrum for a naked singularity formed
in the collapse of a marginally bound inhomogeneous dust cloud and show
that the spectrum is not black body and admits no simple interpretation.
\vfill
\eject

The singularity theorems of Penrose, Hawking and Geroch ensure that,
under fairly general conditions, the collapse of a very massive star will
end in the formation of a singularity.  The theorems do not by themselves
indicate, however, whether the singularity will be covered by an event horizon
or whether it will be visible to  the external observer. If the singularity is
visible to the external observer, it is said to be naked. Naked singularities
are further categorized according to whether or not they are visible to the
asymptotic observer. Those that are visible asymptotically are globally
naked, otherwise they are locally naked. Both varieties have long been considered
undesirable for several reasons and this has given rise to the so-called
``Cosmic Censorship Hypothesis''${}^{[1]}$ (CCH) which essentially banishes them
from the physical universe. While the general consensus seems to be in
favor of the CCH, numerous studies of classical collapse in the literature show
that, on the contrary, naked singularities do indeed form from reasonable initial
data${}^{[2]}$ in classical relativity and low energy string induced gravity, making
it necessary to understand just how nature avoids them, if it does so
at all. Indeed, it is possible that the very processes that nature employs to rid the
universe of naked singularities may have astrophysically observable consequences,
which, in turn, can be exploited to obtain experimental information on the behavior
of quantum fields in extremely curved spacetimes and even of quantum gravity
itself.

If matter does indeed attempt to collapse into a shell-focusing naked
singularity, we can expect its behavior to  differ significantly from that of matter
collapsing into a black hole. This is  because, toward the end stages of the collapse,
not only is a region of high curvature accessible to the external observer, making
quantum effects such as particle production and, eventually, the back reaction of
spacetime particularly important, but also because the causal structure of the
spacetime is vastly different. As a consequence of the exposure of regions of very
high curvature, a feature that appears both within the context of exactly solvable models
of two dimensional dilaton gravity${}^{[3]}$ and of more traditional collapse models in
Einstein gravity${}^{[4,5,6]}$ is  the divergence of the radiation flux from the
singularity on the Cauchy horizon. This effect was first observed in two dimensional
stress-tensor calculations given in ref.[6]. On the contrary, black hole radiation approaches
a steady state near the event horizon and shell-crossing naked singularities, which
have been analyzed by Ford and Parker${}^{[7]}$, have been shown not to produce a large
flux of particles.

In earlier articles${}^{[4,5]}$, we had considered particle production due to the
marginally bound, self similar collapse of inhomogeneous dust in Einstein gravity. The
model admits a globally naked singularity at the center when the mass parameter falls
within a certain range and gives rise to a black hole otherwise. Our analysis
established in two ways that the stress tensor of evaporation grows without limit as the
Cauchy horizon is approached. Naturally, the evaporating cloud cannot radiate away more 
energy than it has, implying that, barring some unexpected effect due to the back reaction 
of spacetime, the Cauchy horizon can safely be expected not to form at all. Thus, the CCH 
(at least in its weak form) would have its origin in the quantum theory, not in the 
classical. Given that the radiation flux is large, particularly toward the end stages of 
the collapse, it is natural to expect that the effects of the flux may be observable. If 
so, what would be its signature? In search of the answer to this question, we will consider
the spectrum of the radiation emitted from the collapsing cloud when the mass parameter is
such that a naked singularity would have been classically formed. As it turns out, the
spectrum is unique and has no simple interpretation in terms of known distributions. Contrary
to the intensity of the radiation flux at infinity which, as we have said before, is due to
the regions of high curvature encountered, the peculiar nature of the spectrum we derive owes
to the causal structure of the spacetime. It is particularly interesting as its
very uniqueness may  serve to distinguish experimentally between naked singularities and
other celestial radiators.

In the following, we have neglected the scattering of waves within the collapsing
cloud and the back reaction of spacetime, confining ourselves to the geometric optics
approximation. Clearly, both of these simplifications will imply that
corrections to the spectrum obtained below are to be expected. These corrections will be
frequency dependent and their importance will depend on how close the collapsing
cloud actually comes to forming the Cauchy horizon. The more important corrections to
the spectrum will arise from the back reaction of the spacetime. In low energy string theory
models of gravity, the time at which the back reaction becomes important is estimated by
evaluating the size of the dilaton coupling constant at the center of the infalling matter.
There is no such simple method in classical relativity, but we might guess that it will become
significant for wavelengths that are smaller than the smallest radius of curvature encountered
though, unfortunately, without the benefit of a full quantum theory of gravity, it is
impossible to estimate precisely the strength or nature of these corrections.

Let us begin by briefly reviewing the classical collapse of a marginally bound inhomogeneous
dust cloud. The matter is described by the stress energy tensor
$$T^\mu_\nu~~ =~~ \epsilon(t,r) \delta^\mu_0 \delta^0_\nu \eqno(1)$$
and the solution of Einstein's equations in comoving coordinates is the Tolman-Bondi${}^{[7]}$
metric
$$ds^2~~ =~~ dt^2~  -~ \tR^{'2}(t,r) dr^2~ -~ \tR(t,r)^2 d\Omega^2 \eqno(2)$$
where
$$\tR~(t,r)~~ =~~ r \left(1~ -~ {3 \over 2} \sqrt{{F(r)} \over {r^3}} t \right )^{2/3}\eqno(3)$$
and $\tR'(t,r)$ is the partial derivative of $\tR(t,r)$ with respect to the coordinate
$r$. The function $F(r)$ is called the mass function. We will consider the special case
of ``self-similar'' collapse which  corresponds to a specific choice of $F(r)$, namely
$F(r) = \lambda r$ where $\lambda$ is a positive constant and is related to the mass
of the collapsing matter. The solution is obviously valid only in the interior of the
cloud, that is up to a fixed value, $r_o$, of the coordinate $r$, giving the total mass as
$2M = \lambda r_o$. Beyond $r_o$ the metric is Schwarzschild and the two regions,
the interior and the exterior, can be made to match smoothly at $r=r_o$. It can be
shown that $\tR(t,r) = 0$ is a curvature singularity which gives the singularity curve
in comoving coordinates as $t = 2r/3\sqrt{\lambda}$.

The solution above has been examined in detail in [4] and here we will only describe the
spacetime. For self-similar collapse, the null coordinates  inside the cloud are given as
$$\eqalign{u~~ &=~~ \left\{\matrix{+r e^{I_-}&~~~ x-\tR' > 0\cr-r e^{I-}&~~~ x-
\tR' < 0}\right.\cr v~~ &=~~ \left\{\matrix{+r e^{I_+}&~~~ x+\tR' > 0\cr-r e^{I_+}
&~~~ x+\tR'<0}\right.\cr}\eqno(4)$$
in terms of $x=t/r$ and the integrals $I_\pm$ defined by
$$I_\pm(t,r)~~ =~~ \int {{dx} \over {x \pm \tR'}}\eqno(5)$$
They have been defined so as to reduce to the usual null coordinates of flat
Minkowski spacetime in the limit as $\lambda \rightarrow 0$. To analyze the
causal structure it is convenient to use the variable $y = \sqrt{\tR(t,r)/r}$
in terms of which the integrals in (5) can be expressed as
$$I_\pm(y)~~ =~~ 9\int {{y^3 dy} \over {f_\pm(y)}}\eqno(6)$$
where $f_\pm(y)$ are the quartic polynomials
$$f_\pm(y)~~ =~~ 3y^4~ \mp~ {{3\sqrt{\lambda}} \over 2}y^3~ -~ 3y~ \mp~
3\sqrt{\lambda} \eqno(7)$$
The center of the cloud can be shown to be just the line $u=v$, whereas
the singularity curve away from the origin is spacelike and given in terms
of the null coordinates of (4) by $v=-cu$ where $c$ is a positive constant. It is
the origin, $u=0=v$, that is of interest, the singularity being globally naked here if and
only if there exists at least one positive real root of the polynomial $f_-(y)$ defined in
(7).${}^{[4,8]}$ Each positive real root of $f_-(y)$ corresponds to an out going null ray that
reaches $\scrip$, whereas each positive real root of the polynomial $f_+(y)$
corresponds to an infalling null ray originating on $\scrim$ and intersecting the
origin. Now it can be shown that $f_-(y)$ admits two positive real roots when the
mass parameter is less that a critical value, $\lambda_c \sim 0.18$, and none
when the mass parameter $\lambda > \lambda_c$, while the polynomial
$f_+(y)$ always admits a pair of real roots, one negative and one positive,
when $\lambda$ is in this range. $\lambda_c$ is thus a critical point. The
singularity at the origin is globally naked for $\lambda < \lambda_c$ and for
$\lambda > \lambda_c$ it is covered. In the latter case, the collapse leads
to the formation of a black hole.

The exterior region is described by the usual Schwarzschild metric
$$ds^2~~ =~~ {{e^{-R(U,V)/\kappa}} \over {R(U,V)}} dU dV~ -~
R^2(U,V) d\Omega^2\eqno(8)$$
where $T,R$ are the Schwarzschild coordinates, $\Omega$ is the solid angle
and $U,V$ are the Kruskal coordinates defined in the usual way by
$$\eqalign{U~~ &=~~ -2\kappa e^{-\tU/2\kappa}\cr V~~ &=~~ 2\kappa
e^{\tV/2\kappa}.\cr}\eqno(9)$$
in which definition $\kappa = 2M$ and  $\tU,\tV$ are the Eddington-Finkelstein
outgoing and incoming null coordinates, $\tU = T - R_*$, $\tV = T+R_*$
respectively. $R_*$ is the tortoise coordinate. The two spacetimes in (8) and
(2) are matched by matching the first and second fundamental forms at the boundary
$r=r_o$.${}^{[7]}$ Comparing the angular parts of the metrics, it is natural to let $R(t) =
\tR(t,r_o)$ at the boundary, whence one easily derives the relationship between
the Schwarzschild null coordinates and the variable $y$ on the boundary,
$$\eqalign{\tU(y)~~ &=~~ -~ {{3r_o} \over {2\sqrt{\lambda}}} y^3~ -~
2 \sqrt{\lambda} r_o y~ -~ r_o y^2 ~ -~  2 \lambda r_o \ln |y/\sqrt{\lambda} - 1|\cr
\tV(y)~~ &=~~ -~ {{3r_o} \over {2\sqrt{\lambda}}} y^3~ -~  2 \sqrt{\lambda}
r_o y~ +~ r_o y^2 ~ +~ 2 \lambda r_o \ln |y/\sqrt{\lambda} + 1|.\cr}\eqno(10)$$
While it is difficult to obtain an analytical relation between the null coordinates
in the interior and exterior regions all along the boundary, it is also not necessary. We
will be interested in relating the two only near the Cauchy horizon. The Penrose diagram
for the classical formation of a naked singularity is shown in figure I below.
\vskip 0.1in
\centerline{\epsfxsize=2.50in \epsfbox{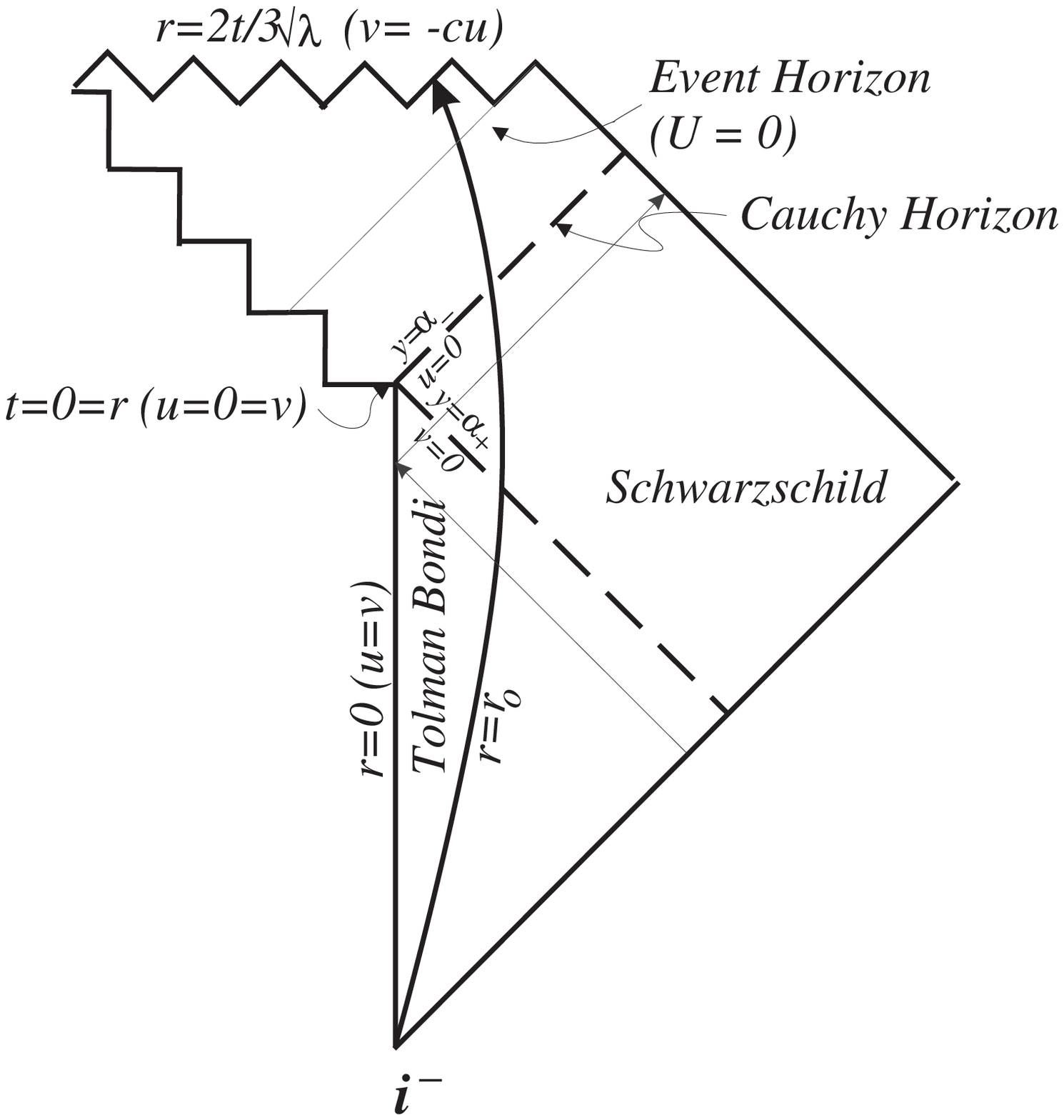}}
\vskip 0.1in
\centerline{\Tenpoint Figure I: Formation of a naked singularity when $\lambda
< \lambda_c$}
\vskip 0.1in

We will consider a massless scalar field, $\phi$, propagating in this background. In the
infinite past, on $\scrim$, the field operator, ${\hat \phi}$ can be expanded unambiguously in
terms a complete, orthonormal family which contains only positive frequencies with
respect to the canonical affine parameter on $\scrim$. This expansion defines the ``in''
vacuum. In the region close to where the Cauchy horizon would classically be formed, the
massless field should also be determined by its data on the Cauchy
horizon and on future null infinity, $\scrip$. Two problems are usually raised about this
program, viz. (a) the absence of a well defined initial value problem to the future of the
intersection of the Cauchy horizon with $\scrip$ implies that there is no justification for
assuming that the totality of $\scrip$ exists, putting in question the completeness of the
mode functions there and (b) because of the singularity at $u=0=v$, the construction of a
complete orthonormal basis set of infalling waves on the Cauchy horizon would be impossible.
The former objection is less serious in that the spacetime to the future of the Cauchy
horizon is the analytic continuation of the spacetime to its past and, in as much as no
singularity actually intersects $\scrip$, there is no {\it a priori} impediment to constructing
a complete basis set on all of $\scrip$. The latter objection appears more serious, but it is
based on the assumption that the Cauchy horizon will actually form. A simple argument shows that
it cannot. In refs. [4,5] the power radiated across future null infinity in this model was
calculated and seen to grow, to leading order, as
$$P(\tU)~~ =~~ ~{1 \over {48\pi}}\left[{{\gamma^2-1}\over {(\tU_o - \tU)^2}}
\right] \eqno(11)$$
where $\tU = \tU_o$ is the Cauchy horizon. There is every indication that the result
is generic (i.e., independent of the actual collapse model considered as long as it leads
classically to the formation of a naked singularity) and it was first suggested by two dimensional
stress-tensor calculations in ref.[6]. Above, $\gamma$ is a parameter${}^{[4]}$ that depends in
a complicated way on the mass parameter, $\lambda$. It approaches unity when $\lambda \rightarrow
0$ and increases, approaching $\sim 1.53$ when $\lambda \rightarrow \lambda_c$. The total
radiated energy is the integrated flux and, because the collapsing cloud cannot radiate more energy
than it possesses, an upper limit being provided by the total mass, $M$ of the cloud, one
may estimate the retarded time, $\tU'$, at which the cloud will have given up all its energy
to the Hawking radiation. It is
$$\tU'~~ \sim~~ \tU_o~ -~ {{\gamma^2-1} \over {12\pi M}}~~ <~~ \tU_o.
\eqno(12)$$
For a sufficiently large total mass, $M$, the collapsing cloud can approach arbitrarily close
to forming the Cauchy horizon, but it will {\it not} actually do so. Note that
we have neglected the back reaction in the estimate above. Naturally, this will become important
toward the end stages of the evaporation, before the cloud has radiated away completely,
and at this point the spacetime will begin to change significantly. This change will be such as
to prevent the Cauchy horizon from forming. Therefore, for a complete, orthonormal set of solutions
to the wave equation which are purely infalling in the semi-classical picture, it is more reasonable
to consider a null surface parallel to and in the past of the classical Cauchy horizon. Such a
choice would not be plagued by the difficulty in (b) above. Moreover, the precise choice of surface
and functions is irrelevant because the functions in this basis play the role of spectator and
do not influence the final result.

Let us imagine that a null ray $\tU = $ const., when traced backwards is found to
originate in the infalling null ray $\tV = {\cal G}(\tU)$. From the general theory,${}^{[10]}$
one knows that the number distribution of Minkowski particles observed on $\scrip$ is simply given
by the Bogoliubov coefficient
$$\beta(\w',\w)~~ =~~ \int_{-\infty}^\infty {{d\tU} \over {4\pi\sqrt{\w\w'}}}
e^{-i\w\tU} e^{-i\w'{\cal G}(\tU)} \eqno(13)$$
(where the integral is performed over all of $\scrip$) as
$${}_M \langle 0| N(\w)|0\rangle_M~~ =~~ \int_0^\infty d\w'
|\beta(\w',\w)|^2 \eqno(14)$$
where $|0\rangle_M$ is the Minkowski vacuum.

When a globally naked singularity is due to be formed in the classical theory, the putative
Cauchy horizon lies in the retarded past of the event horizon at some finite value, $\tU_o$,
of $\tU$ and, as we mentioned before, no radiation can be expected beyond this point as the
cloud will have already given up all of its energy. This is consistent with the semi-classical
picture in which it is easy to see that no incoming rays on $\scrim$ may turn into outgoing rays
in the retarded future of the Cauchy horizon. As $\lambda < \lambda_c$, $f_-(y)$ admits two
positive (real) roots each of which corresponds to a null outgoing ray from the origin. Let
$\alpha_i$ represent the real roots of $f_-(y)$. Because $y=\alpha_i$ implies that
$t = 2r(1-\alpha_i^3)/3\sqrt{\lambda}$, we take the larger of these two roots
as the one that gives the earliest null ray from $u=0=v$ and call it
$\alpha_-$. The Cauchy horizon is therefore given by ($y=\alpha_-$) $u=0$
in the interior and
$$\tU_o~~ =~~ {{3r_o} \over {2\sqrt{\lambda}}} \alpha_-^3~ -~  2 \sqrt{\lambda}
r_o \alpha_-~ -~ r_o \alpha_-^2 ~ -~  2 \lambda r_o \ln |\alpha_-/\sqrt{\lambda}
- 1| \eqno(15)$$
in the exterior. As before, consider a ray that leaves the origin near the
Cauchy horizon and in its retarded past. Such a ray will intersect the
boundary of the cloud at some $u =$ const. such that $y$ is close to
$\alpha_-$, say $y=\alpha_-~ +~ \ty_-$. For small $\ty_-$,
$$I_-(y)~~ \sim~~ \gamma_- \ln \ty_-~ +~ {\cal O}(\ty_-) \eqno(16)$$
where $\gamma_- = 3\alpha_-^3/f'_-(\alpha_-)$, the prime denoting a
derivative w.r.t. $y$. This gives
$$\ty_-~~ =~~ y - \alpha_-~~ =~~ \left(-{u \over {r_o}}\right)^{1/\gamma_-}
\eqno(17)$$
Up to linear order in $\ty_-$ one therefore finds that
$$\tU~~ \sim~~ U_o~  +~ \Gamma_- \ty_-~~ =~~ U_o~ +~ \Gamma_- \left(
-{u \over {r_o}}\right)^{1/\gamma_-} \eqno(18)$$
where $\Gamma_-(\alpha_-) = - 2r_o \alpha_-^3/\sqrt{\lambda}(\alpha_-
- \sqrt{\lambda})$. For $\lambda < \lambda_c$, $\Gamma_-$ is negative
and monotonically increasing as a function of $\lambda$.  Applying the
same reasoning to the infalling ray, which in the interior is given by
$v=0$, one finds that any infalling ray close to $v=0$ and in its
advanced past will intersect the boundary at $y=\alpha_+ + \ty_+$,
where $\alpha_+$ is the positive (real) root of the polynomial $f_+(y)$,
such that
$$\ty_+~~ =~~ y - \alpha_+~~ =~~ \left(-{v \over {r_o}}\right)^{1/\gamma_+}
\eqno(19)$$
where $\gamma_+ = 3\alpha_+^3/f'_+(\alpha_+)$.  Thus one has
$$\tV~~ \sim~~ V_o~ +~ \Gamma_+ \left(-{v \over {r_o}}\right)^{1/\gamma_+}
\eqno(20)$$
where $\Gamma_+(\alpha_+) = - 2r_o \alpha_+^3/\sqrt{\lambda}(\alpha_+
+ \sqrt{\lambda})$ and $\tV_o = \tV(\alpha_+)$ (as given in (10). Combining
(18) and (20) after ``reflecting'' at the center ($u=v$) gives the relationship
between $\tV$ and $\tU$ as
$$\tV~~ =~~ {\cal G}(\tU)~~ =~~A~ -~ B (\tU_o - \tU)^{\gamma}~~~~~  \tU < \tU_o
\eqno(21)$$
where $\gamma = \gamma_-/\gamma_+$ and there are no incoming rays
corresponding to outgoing rays in the retarded future of the Cauchy horizon.
$A=U_o$ and $B (>0)$ are irrelevant constants that can be evaluated explicitly.
When $\lambda = 0$ (there is no infalling matter), one finds that $\gamma = 1 = B$
so that ${\cal G}(\tU) = \tU$ as expected. Thus, the Bogoliubov coefficient,
$\beta(\w',\w)$, when $\gamma > 1$ is
$$\beta(\w',\w)~~ =~~ {1 \over {2\pi}}\sqrt{\w\over \w'} e^{-i\w' A} \int_{-\infty}^{\tU_o}
d\tU e^{-i\w\tU} e^{i\w' B(\tU_o - \tU)^{\gamma}}\eqno(22)$$
Changing variables to $z = (\tU_o -\tU)$, one has
$$\beta(\w',\w)~~ =~~ {1 \over {2\pi}} \sqrt{\w \over \w'} e^{-i\w'A - i\w \tU_o}
\int_0^\infty dz e^{i\w z} e^{i \w' B z^\gamma} \eqno(23)$$
which gives
$$|\beta(\w',\w)|^2~~ =~~ {1 \over {4\pi^2\w\w'}} |\sum_{k=0}^\infty
{{(iB \w' \w^{-\gamma} e^{i\pi\gamma/2})^k} \over {k!}}  \Gamma (k\gamma+1)|^2
\eqno(24)$$
The radiation from the naked singularity is clearly not black body radiation, as it is
for a black hole. Indeed we are unable to find any analogy between (31) and
the standard distributions that arise in particle physics.

The expression above is useful to analyze the high frequency limit ($B\w'(\w)^{-\gamma}
\rightarrow 0)$ limit of the spectrum, for in this limit it is
sufficient to consider only the first term in the series. Integration over $\w'$ then
yields the familiar logarithmic divergence in the high frequency region. The
alternative expression
$$|\beta(\w',\w)|^2~~ =~~ {\w \over {4\pi^2  \gamma \w' (\w' B)^{2/\gamma}}}
|\sum_{k=0}^\infty {({i\w(\w' B)^{-1/\gamma}e^{-i\pi /2\gamma})^k} \over
{k!}} \Gamma({{k+1} \over \gamma})|^2\eqno(25)$$
serves to analyze its low frequency ($B\w'(\w)^{-\gamma} \rightarrow \infty$)
behavior. Integration over $\w'$ in this limit shows a power law divergence in the
infrared. This divergence is associated with the fact that there are an infinite number
of quanta in each mode on $\scrip$. As the collapsing ball produces a steady flux of
radiation to $\scrip$, the net flux for all time is infinite.  The difference between the
divergence in the low and high frequency regimes may be associated with the red-shifting
of modes in the proximity of the putative Cauchy horizon. Nevertheless, $|\beta(\w',\w)|^2$
is seen to be well behaved as a function of $\w$. It approaches $\w$ in the infrared
and $1/\w$ in the ultraviolet regions.

Both the low frequency behavior of the black hole and the naked singularity as well as
their high frequency behavior are seen to differ significantly. At low frequencies the
intensity of the radiation from a naked singularity drops off as $\w$, whereas for a
black hole it behaves as $1/\w$. In the same way, at high frequencies, the radiation
spectrum of a naked singularity drops off in intensity as $1/\omega$ whereas the black
hole spectrum drops off much faster, as $e^{-\beta_H \omega}$ where $\beta_H$ is the
inverse Hawking temperature. Naked singularities prefer to radiate at higher frequencies.

In this paper we have described a potentially important astrophysical problem, namely that of
a gravitational collapse that classically leads to the formation of a naked singularity. We
have argued that the presence of strong gravitational fields accessible to the asymptotic
observer imply that the naked singularity is likely not to form because the matter energy of the
collapsing cloud will disperse as Hawking radiation before the arrival of the Cauchy horizon.
This radiation will be intense and possibly observable asymptotically. We have also shown that
the unique causal structure associated with a global naked singularity leads to a unique spectrum.

In our treatment, we have made no attempt to include the back reaction of the spacetime though
it may be argued that it is precisely the back reaction that will dominate the evolution of the
system in the final stages of collapse. The back reaction is a serious issue and should be treated
satisfactorily at some point. As far as we know, only string theory or its generalizations
provide us with consistent quantum theories of gravity, so it would appear that it can be
satisfactorily studied only in the context of string gravity. Yet, in the few known
exactly solvable models of collapse in string theory,${}^{[3,13]}$ the findings
have not so far contradicted the general conclusions which have been arrived at by the
semi-classical approach. It is possible therefore that the above description is reasonably
accurate.

The subject of gravitational collapse in Einstein gravity is rich and cannot be ignored. The purpose
of this work is to initiate a discussion on the subject and to attempt to find observational signatures
that distinguish between matter collapsing to form a black hole and matter collapsing to form a naked
singularity, even though that end state may never be achieved. (Black hole studies, for instance, are
very useful even though the end state of the black hole is not known.) Our hope is that whatever
information one can extract from models of collapse in Einstein gravity may later be incorporated into
a better understanding of the role of the spacetime from string (or, for example, $M-$) theory.

To conclude, we venture to summarize in terms of a Penrose diagram (figure II) what we believe
the spacetime in the fully quantized theory should look like, based upon the understanding
that has been obtained from the semi-classical findings related above.
\vskip 0.1in
\centerline{\epsfxsize=1.5in \epsfbox{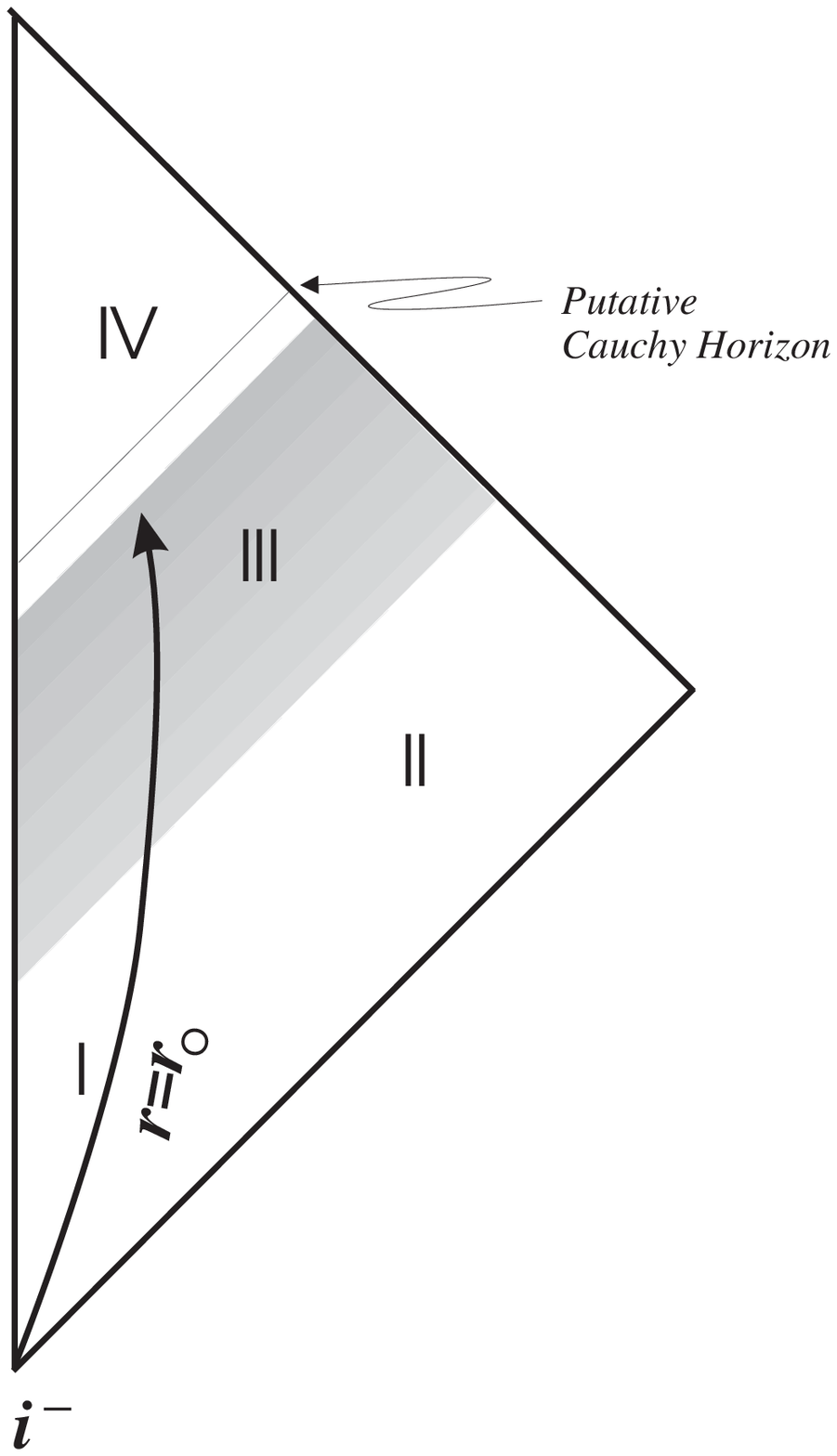}}
\vskip 0.1in
{\Tenpoint \narrower\narrower\narrower\smallskip\noindent{\bf Figure II:}
Conjectured structure of the spacetime in the full quantum theory of a cloud attempting to
form a classical naked singularity\smallskip}
\vskip 0.2in
In Figure II, region I is described by the Tolman-Bondi metric given in eq.(2). Region II is
the Schwarzschild exterior and, in region III, Hawking radiation becomes significant so that
energy streams out of the cloud toward infinity. Strictly speaking, region III also
contains collapsing matter, but the radiation dominates. As such it is probably acceptable to
model this patch by the Vaidya metric. It must, of course, be matched to regions I and II in
the usual way. It is along $\scrip$ in this portion of spacetime that the spectrum obtained above
should be a good approximation until the end stages when the back reaction kicks in. Notice that
the putative Cauchy horizon is to the future of this region. It falls in a patch (region IV) that
is Minkowskian because, presumably, all the energy has by now escaped to infinity. This scenario
will be discussed and the limitations of the approximation above will be systematically treated
more fully elsewhere.
\vskip 0.1in

\noindent{\bf Acknowledgements:}

\noindent We acknowledge the partial support of the {\it Junta Nacional de Investiga\c{c}\~ao
Cient\'\i fica e Tecnol\'ogica} (JNICT) Portugal, under contract number CERN/S/FAE/1172/97 and
the partial support of NATO, under contract number CRG 920096. L.W. acknowledges the partial
support of the U. S. Department of Energy under contract number DOE-FG02-84ER40153.
\vfill\eject

\noindent{\bf References:}

{\item{[1]}}R. Penrose, Riv. Nuovo Cimento {\bf 1} (1969) 252; in {\it General Relativity, An Einstein
Centenary Survey}, ed. S. W. Hawking and W. Israel, Cambridge Univ. Press, Cambridge, London (1979)
581. In its original form, the Cosmic Censorship Hypothesis (CCH) essentially states that {\it
no physically realistic collapse, evolving from a well posed initial data set and satisfying the
dominant energy condition, results in a singularity in the causal past of null infinity}. There is
also a strong version of the CCH which states that {\it no physically realistic collapse leads to a
locally timelike singularity}.

{\item{[2]}}P. S. Joshi, {\it Global Aspects in Gravitation and Cosmology}, Clarendon Press, Oxford,
(1993).

{\item{[3]}}Cenalo Vaz and Louis Witten, Phys. Letts. {\bf B325} (1994) 27; Class. Quant.
Grav. {\bf 12} (1995) 1; {\it ibid.} {\bf 13} (1996) L59; Nucl. Phys. {\bf B487} (1997)
409.

{\item{[4]}}S. Barve, T.P. Singh, Cenalo Vaz and Louis Witten, ``Particle Creation in the
Marginally Bound Self-Similar Collapse of Dust'', gr-qc/9802035, Nucl.
Phys. {\bf B} {\it in press}.

{\item{[5]}}S. Barve, T.P. Singh, Cenalo Vaz and Louis Witten,
``The Quantum Stress Tensor in Self-Similar Spherical Dust
Collapse'', gr-qc/9805095, Phys. Rev. {\bf D} {\it in press}.

{\item{[6]}}W. A. Hiscock, L. G. Williams and D. M. Eardley, Phys. Rev. {\bf D26} (1982) 751.

{\item{[7]}}L. H. Ford and Leonard Parker, Phys. Rev. {\bf D17} (1978) 1485.

{\item{[8]}}R. C. Tolman (1934) Proc. Nat. Acad. Sci. USA {\bf 20} 169; H. Bondi (1947) Mon. Not.
Astron. Soc. {\bf 107} 410.

{\item{[9]}}P.S. Joshi and T.P. Singh, Phys. Rev. {\bf 51} (1995) 6778, and references therein.

{\item{[10]}}See, for example, N. D. Birrel and P.C.W. Davies, {\it Quantum Fields in
Curved Space}, Cambridge Monographs in Math. Phys., Cambridge University Pres, London
(1982).

{\item{[11]}}S. W. Hawking, Comm. Math. Phys. {\bf 43} (1975) 199.

{\item{[12]}}D. Page, Phys. Rev. {\bf D13} (1976)  198.

{\item{[13]}}C. Callan, S. B. Giddings, J. Harvey and A. Strominger, Phys. Rev. {\bf D45} (1992) 1005;
S. de Alwis, Phys. Lett. {\bf B289} (1992) 282; A. Bilal, C. Callan, Nucl. Phys. {\bf B394} (1993) 73;
Sukanta Bose, L. Parker and Yoav Peleg, Phys. Rev. {\bf D52} (1995) 3512; {\it ibid}
Phys. Rev {\bf D53} (1996) 5708; {\it ibid}, Phys. Rev. {\bf D53} (1996) 7089; {\it ibid}, Phys. Rev.
Letts. {\bf 76} (1996) 861; A. Fabri and J. G. Russo, Phys. Rev. {\bf D53} (1996) 6995.

\end